\documentclass{icrc2009}

\usepackage{graphicx}   
\usepackage[caption=false]{caption}    
\usepackage[font=footnotesize]{subfig} 
\usepackage{fixltx2e}
\usepackage{url}

\newcommand{\shorttitle}[1]%
{\markboth{Proceedings of the 31\MakeLowercase{$^{st}$} ICRC, {\L}\'{o}d\'{z} 2009}{#1} }
\newcommand{\etal}{\MakeLowercase{\textit{et al. }}} 

\begin{document}
\title{HAWC Timing Calibration}

\author{\IEEEauthorblockN{Petra H\"untemeyer\IEEEauthorrefmark{1},
			  John A.~J.~Matthews\IEEEauthorrefmark{2},
                          Brenda Dingus\IEEEauthorrefmark{3}, 
                          for the HAWC Collaboration}
                            \\
\IEEEauthorblockA{\IEEEauthorrefmark{1}Michigan Technological University, Houghton, MI}
\IEEEauthorblockA{\IEEEauthorrefmark{2}University of New Mexico, Albuquerque, NM}
\IEEEauthorblockA{\IEEEauthorrefmark{3}Los Alamos National Laboratory, Los Alamos, NM}
}

\shorttitle{Petra H\"untemeyer \etal HAWC Timing Calibration}
\maketitle

\begin{abstract}
The High-Altitude Water Cherenkov (HAWC) Experiment is a second-generation
high-sensitivity gamma-ray and cosmic-ray detector that builds on the
experience and technology of the Milagro observatory. Like Milagro, HAWC
utilizes the water Cherenkov technique to measure extensive air
showers. Instead of a pond filled with water (as in Milagro) an array of
closely packed water tanks is used. The event direction will be
reconstructed using the times when the PMTs in each tank are 
triggered. Therefore, the timing calibration will be crucial for reaching  
an angular resolution as low as 0.25 degrees. We propose to use a laser 
calibration system, patterned after the calibration system in
Milagro\cite{milagro-laser}.  Like Milagro, the HAWC optical calibration system
will use $\sim$1~ns laser light pulses.  Unlike Milagro, the PMTs are 
optically isolated and require their own optical fiber calibration. 
For HAWC the laser light pulses will be directed through 
a series of optical fan-outs and fibers to illuminate the PMTs in approximately
one half of the tanks on any given pulse.  Time slewing corrections will 
be made using neutral-density filters to control the light intensity 
over 4 orders of magnitude.
This system is envisioned to run continuously at a low rate and will be
controlled remotely. In this paper, we present the design of the
calibration system and first measurements of its performance.

 \end{abstract}

 
\section{Introduction}
The High-Altitude Water Cherenkov (HAWC) Experiment will be built at 4100 
meters on a plateau below the highest mountain in Mexico, the Pico de Orizaba.
It will be 10 to 15 times more sensitive than its predecessor, the 
Milagro Observatory\cite{milagro-reference}. 
Like Milagro, HAWC will survey the sky (from its location at
19$^{\rm o}$ north latitude) continuously measuring extensive air showers 
generated by cosmic and gamma rays, leading to a large field-of-view of 
$\sim$2~sr. The physics goal of HAWC is to make major progress in studying 
the origin of of cosmic rays. This can be achieved by detecting the
highest energy
gamma-ray sources ($\sim$100 TEV) and by measuring and mapping the galactic 
diffuse gamma-ray emission from 1 TeV to 100 TeV. Because of its $\sim 100$\%
duty cycle, HAWC is particularly well suited to search for transients,
new galactic and extragalactic sources of VHE gamma-radiation, 
and for small- and 
large-scale anisotropies in the cosmic radiation in an unbiased sky survey.
In order to meet these physics goals, good angular resolution is essential. 
With HAWC we aim for an angular resolution between 0.25$^{\circ}$ 
and 0.55$^{\circ}$, where 0.55$^{\circ}$ is representative of low energies and
0.25$^{\circ}$ of high energies.  As in
Milagro the shower direction will be reconstructed using the relative
times at which each of the PMTs is hit. In this paper, we will describe 
the design of the optical calibration system that will be used to 
monitor the relative time responses of each PMT in the HAWC detector.
 
\section{The HAWC Detector}

\begin{figure}[!t]
  \centering
  \includegraphics[width=2.5in]{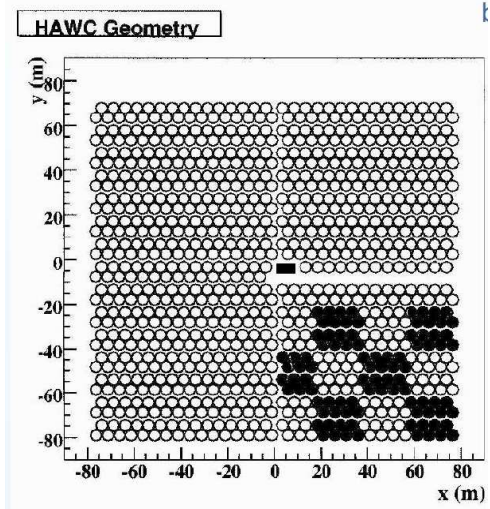}
  \caption{A sketch of one proposed layouts for the tanks of the HAWC 
detector. The rectangle in the center represents the counting house;
this is also the site of the calibration light sources. 
The black and white pattern of the tanks (in the lower-right of the figure)
is explained in the text.}
  \label{fig-array}
 \end{figure}

The HAWC Observatory is described in more detail elsewhere in these 
Proceedings~\cite{jordan-magda}.
It will re-use the 900 8" PMTs\cite{hamamatsu-8inch} 
from Milagro. The PMTs will
be deployed at a depth of  $\sim$4 m of water in separate tanks. 
The tanks will be 
closely packed in an array covering an area of approximately 
150m x 150m. Depending on the detailed design, 
more than 60\% of this area will be covered with water.
Two alternative designs are under consideration: 
one with 900 individual rotomolded
water tanks of $\sim 4$m diameter holding one PMT; a second with 7.3m 
diameter tanks (which are a combination of corrugated galvanized steel
walls and a water-tight bladder) holding three PMTs. Each tank will
have one or more 8" baffled, upward-facing PMT(s) anchored to the bottom. 
Figure~\ref{fig-array} shows a sketch of the proposed 900 tank deployment 
pattern. 

\section{The Calibration System}

The primary goal of the optical calibration system is to monitor the time 
stability of the PMTs and related readout/digitization electronics. 
To achieve an angular resolution of $<$0.55$^{\circ}$ 
it is not only necessary to know the position of each PMT to only a few cm, 
but also to know the relative times at which each PMT in the array
has been hit to an uncertainty of $\le 1$ns.  Furthermore the 
channel to channel timing drifts and the slewing must be monitored and 
corrected for accordingly\cite{milagro-calibration}.
The components of the HAWC optical calibration system are chosen to meet 
this goal.  This includes {\it e.g.} the $<1$ns pulse-width laser light 
source\cite{teem} and the use of equal length, duplex optical fibers
on all fiber paths.
Furthermore to monitor the time slewing and to measure the
{\it photo-electron calibration} of each (PMT) 
channel\cite{milagro-calibration}, filter wheels
at the laser light source will allow control of the light pulses over
a dynamic range from $\sim 0.1$ PE to $> 1000$ PEs.

For reasons of practicality and redundancy, the tank PMTs are divided into 
two groups: denoted {\it black} and {\it white} for optical calibration.  
This is sketched in Figure~\ref{fig-array} where $\sim 16$ physically 
nearby PMTs form a black (or white) {\it square}.  The pattern of {\it squares}
is similar to the pattern on a checker(chess) board.  A possible pattern
of {\it squares} is shown (for the lower-right quadrant of the array) in
Figure~\ref{fig-array}.  The calibration
design foresees two separate laser light delivery systems. 
One laser plus optical
distribution system will be used
to pulse all of the black tanks, a second system will pulse all of the white
tanks. The two laser systems and associating monitoring instrumentation will be
located in a temperature controlled enclosure at the center of
the HAWC array (shown as black rectangle in the center of the array in
Figure~\ref{fig-array}).

\begin{figure}[!t]
  \centering
  \includegraphics[width=3.0in]{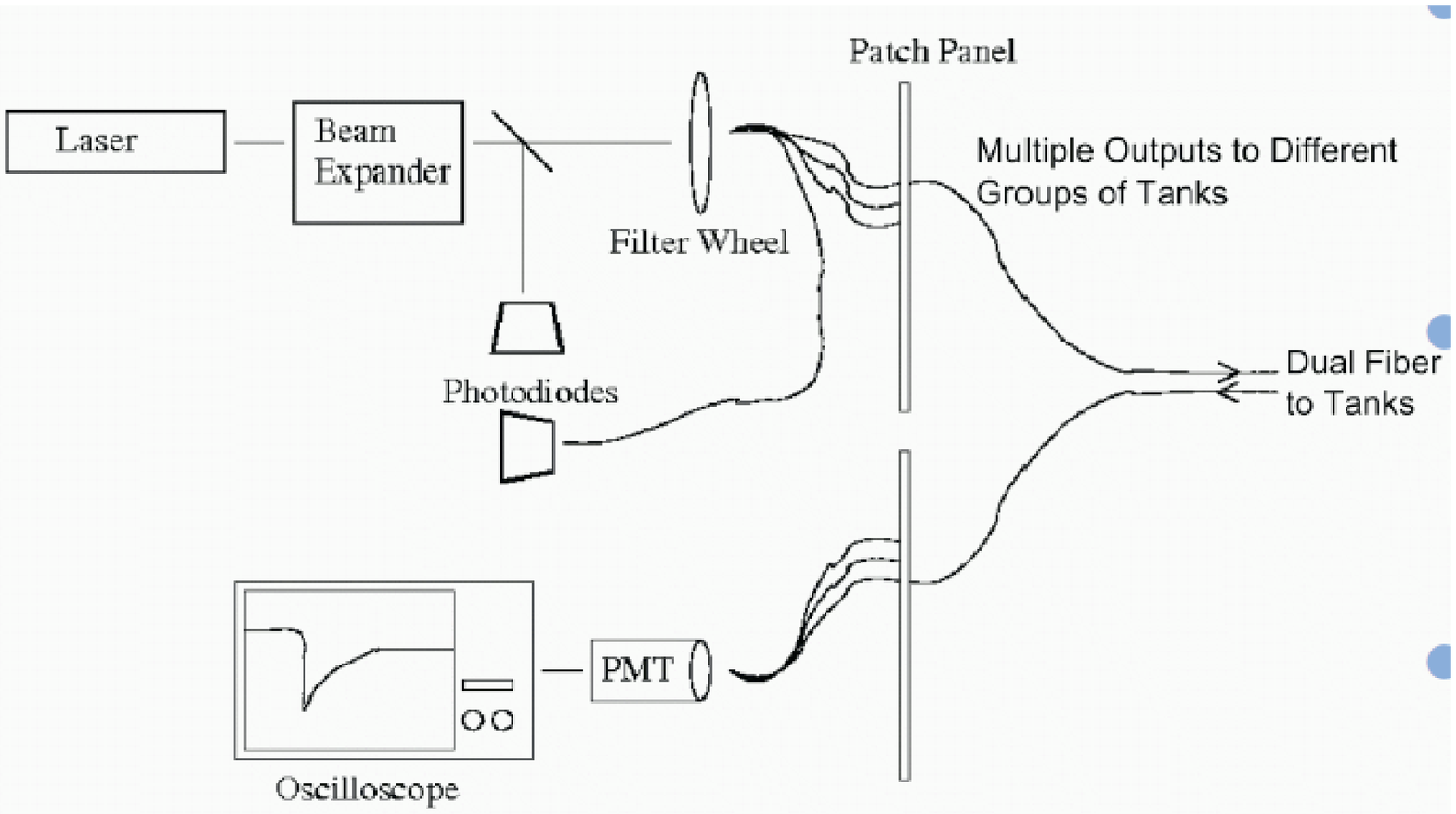}
  \caption{Sketch of one of two HAWC {\it light sources} for the optical
   timing calibration system.  The {\it light sources} will be located at the
   center of the array; see Figure~\ref{fig-array}.  See text for details.}
  \label{fig-lightsource}
 \end{figure}

\begin{figure}[!h]
  \centering
  \includegraphics[width=3.0in]{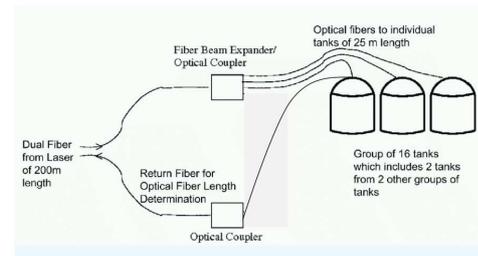}
  \caption{Sketch of one of approximately 60 HAWC {\it field distributions} 
  of the laser light for the optical timing calibration system. Each of the
  {\it field distributions} will be located near the center of the
  black (or white) {\it squares}.  The {\it field distributions} further
  split the laser light pulses to the fibers that then go to each PMT.
  Additionally one duplex fiber (per {\it square}) will be run 
  to one (representative) tank and the signal looped-back to be monitored
  for end-to-end timing at the light source; see Figure~\ref{fig-lightsource}.}
  \label{fig-field}
 \end{figure}

Figure~\ref{fig-lightsource} shows a sketch of one of 
the proposed light sources and its associating instrumentation. The 
light from the Teem Photonics PowerChip laser\cite{teem} 
will be sent through a beam expander\cite{thorlabs-beamexpander} into 
a 1:37 optical splitter\cite{romack} at the calibration light source.
The 1:37 optical splitter will fan-out the laser light
pulses into the $^{>}_{\sim}$ 30 paths leading to the centers of all black 
(or white) {\it squares} of tanks.  To connect to the approximate center
of each {\it square} we will use same-length, $\sim$200m long, 
$62.5\mu$m/125~$\mu$m graded index duplex fibers.  Finally at each {\it square}
the light is split again to be routed on $\sim$25m fibers to each tank
(see Figure~\ref{fig-field}) where the fibers will be terminated with
a simple diffuser positioned at a fixed distance from each tank PMT.
Additionally there will be one loop-back duplex
fiber for monitoring (double) the optical fiber light transmission time
for a representative light path for each {\it square}, and $\sim 1$ 
fiber to a tank in the adjacent {\it square} 
to link the timing of black and white optical calibrations.

\vspace{0.2in}

\begin{figure}[!t]
  \centering
  \includegraphics[angle=-90,width=2.5in]{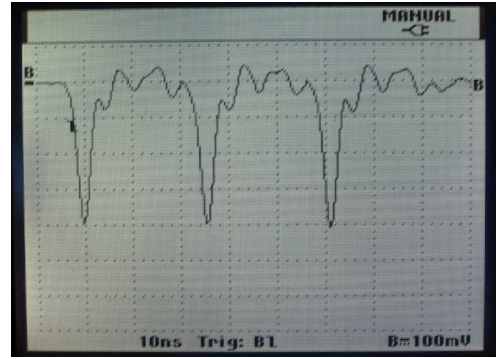}
  \caption{The observed return-light time distribution from three 
  separate calibration light paths.
  The 200 Mhz oscilloscope has somewhat limited bandwidth for the fast
  $\sim 1$ns light pulses.  The relative time delays between the three
  light paths is the result of the addition of 5m and 10m of (additional) 
  optical
  fiber to two of the light paths.  The near equality of the signals shows
  the good uniformity of the light coupling into the 1:37 output 
  and 1:19 return optical fan-outs/fan-ins.}
  \label{fig-scope}
 \end{figure}

The $\sim$200m long, duplex fibers naturally
provide an outgoing and a return light path.  The outgoing light path is
needed to deliver the laser light pulses to the center of each {\it square}.
The return light path is used to monitor the light pulses by returning 
one representative signal; see Figure~\ref{fig-field}.  In summary
the outgoing light follows the path from: the laser, into the 1:37 splitter,
coupled through a patch panel (at the light source 
see Figure~\ref{fig-lightsource}),
into the $\sim$200m long fiber to the {\it field distribution} 
at the center of each {\it square} where 
the light is split again to be routed on $\sim$25m fibers to each tank.
The return light path will be exactly double the total optical fiber light path
and is used to return a representative light signal from each {\it square};
see Figure~\ref{fig-field}.   Again for both redundancy and
practicality reasons, the return light pulses are merged using
four 19:1 optical fan-ins\cite{romack}, one for each quadrant of the array;
see Figure~\ref{fig-lightsource}.  The return light will then be monitored
with a small photo-tube\cite{hamamatsu}.  Up to small timing differences,
all return signals will arrive at the same time.  Thus to distinguish
each light path, we are considering adding optical delay fibers, in
increments of $\sim 5$m of fiber delay, between the return light fibers from
the {\it squares} and the 1:19 optical fan-ins.  In this way each light
path can be easily monitored.  An example, showing the return light from three
separate calibration loop-back signals, 
is shown in Figure~\ref{fig-scope}.  The relative
time delay from 5m optical fiber is 25.6ns.

Assuming a relative timing delay of $\geq$5m of optical fiber,
the distinct light signals observed in the initial proof of principle
study (Figure~\ref{fig-scope}) suggests that it should be
straight forward to monitor all light paths to a precision
of $\sim 1$ns using {\it e.g.} 
a 4-channel digital oscilloscope digitize the return light pulses
for each of the optical calibration light paths.

\section{Tests of the Calibration System}

Several laboratory tests of the calibration system have already been performed.
These have focused on proof of principle tests: {\it e.g.} how uniform are
the light signals in the 37 output fibers of the 1:37 optical splitter
at the light source, or the light signals in the 19 optical fibers of 
the 1:19 optical splitter at the {\it field distribution} at 
each {\it square}.   The other critical issue is the expected light 
at the individual tank PMTs and whether it can result in $>1000$ PEs 
(the maximum signal needed to monitor the PMT/readout time slewing 
corrections).

To achieve uniform fiber-to-fiber illumination with any 1:n splitter
requires that the size of the light spot is significantly larger than
the size of the fiber bundle.  Laser beams can be increased in
size easily using commercial beam expanders.  The 
balance
is then between
fiber-to-fiber uniformity and the signal size in each fiber: more
uniformity comes at the cost of decreased signal size.  First measurements
with the JDS Uniphase Power Chip laser\cite{milagro-laser} find that
expanding the beam at the laser by a factor 
of $\sim 5\times$ results in a rather uniform illumination 
of the 1:37 optical fiber splitter\cite{romack}.  
A LaserProbe RM6600A radiometer with a RjP-465 silicon 
sensor\cite{laserprobe} has been used to measure the laser intensity
and the signals into the optical fibers.  As an example, with a $5\times$
beam expander. typical fiber signals were $\sim 15$nJ ($\pm$25\%) 
per pulse, while the measured laser energy was $\sim 22\mu$J.

In tests of the {\it field distribution},
an aspheric lens with a focal length of 4.5mm\cite{thorlabs-lens} 
was used to create a 
collimated beam from the light source fiber into the 1:19 optical splitter. 
However it was found that the (light source) fiber speckle pattern
resulted in a non-uniform light intensity within the collimated beam.
One solution was to add a simple ground glass diffuser\cite{thorlabs-diffuser} 
between the lens and the 1:19 splitter.  While this decreased the
the light intensity into the individual fibers of the 1:19 splitter,
the net impact is on the design and positioning of the tank diffuser (next
paragraph).  An alternate possibility (yet to be studied)
is to reduce the amount of speckle by randomizing the 
polarization of the laser light at the source.

The design of the tank light diffuser, mounted on the end of the 
optical fiber(s) in each tank, has not been finalized.  However 
estimates of the light intensity at the HAWC tanks have been made
by including all elements of the HAWC calibration up to (but not including)
the final tank optical diffusers.  
The resulting {\it table top} mock-up of the HAWC calibration system 
included: the (existing) PowerChip laser, $5\times$ beam expander,
1:37 fiber splitter, 2 x 150m of $62.5\mu$m/$125\mu$m optical fiber 
distribution, and the {\it field distribution} optics (including the 
ground glass diffuser and 1:19 fiber splitter).
Including a $\sim 8$\% quantum efficiency for the 8'' Hamamatsu 
PMTs\cite{hamamatsu-8inch}
at the calibration wavelength of 532nm, we expect $>1000$ PEs in the PMTs 
as long as the light coupling efficiency is $>1$\% between 
the calibration optical fiber (diffuser) and the 8'' PMTs.
This argues for mounting the tank optical diffusers 
close to the PMTs.  This choice also minimizes possible PMT to PMT 
laser light time arrival differences from any light path length differences 
in the tanks.

\section{Conclusion}

A {\it table-top} realization of the proposed HAWC optical calibration system
was set up to evaluate in detail the timing calibration of the HAWC detector.
The components of the calibration system have been tested for their 
performance, in particular their timing stability and the expected light 
intensity delivered to the individual tanks.  No obstacles have 
been found. Future efforts will focus on optimizing the light yield (and 
fiber to fiber uniformity) as well as the design of the tank diffuser and
tank diffuser mounting plan.

\end{document}